**Planning sustainable carbon neutrality pathways: accounting challenges experienced by organizations and solutions from industrial ecology**

A. de Bortoli[1,2*], Anders Bjørn[3], François Saunier[1], Manuele Margni[1,4]

1 CIRAIG, École Polytechnique de Montréal, P.O. Box 6079, Montréal, Québec, H3C 3A7, Canada

2 LVMT, Ecole des Ponts ParisTech, Université Gustave Eiffel, 5 boulevard Descartes — Champs-sur-Marne — 77454 Marne-la-Vallée Cedex2, France

3 Department of Management, John Molson School of Business, Concordia University, Montréal, Quebec, Canada.

4 HES-SO, University of Applied Sciences and Arts Western Switzerland, Institute of Sustainable Energy, Sion, Wallis, Switzerland

*Corresponding author: Anne de Bortoli, CIRAIG, École Polytechnique de Montréal, P.O. Box 6079, Montréal, Québec, H3C 3A7, Canada. e-mail : anne.debortoli@polymtl.ca

Abstract

*Purpose:* Planning a transition towards sustainable carbon neutrality at the organization level raises several accounting challenges. This paper aims to shed light on key challenges, highlight answers from current accounting standards and guidance, point out potential inconsistencies or limits, and outline potential solutions from the industrial ecology community through systemic environmental assessment tools, such as life cycle assessment (LCA) and environmentally-extended input-output (EEIO) analysis.

*Method:* The study is based on the accounting difficulties related to GHG emissions as well as other sustainability concerns (environmental, social and financial), reported to the authors by multiple organizations in developing carbon neutrality plans. The study then draws on a literature review of carbon neutrality-related standards and guidelines, as well as industrial ecology studies, to identify answers to these reported challenges.

*Results and discussion*: We propose a "Measure-Reduce-Neutralize-Control" sequence allowing organizations to plan their sustainable net-zero strategy, and discuss 24 accounting challenges occurring within this sequence. We then outline ways forward for organizations planning their carbon neutrality trajectory — pointing to existing

resources —, and for guidelines providers and the industrial ecology communities to address current limitations in the development of future accounting methods and guidelines. Overarching solutions to many accounting issues are to develop comprehensive, open-source, and high-quality life cycle inventory databases, to enable improved dynamic assessments and prospective LCA through integrated assessment models, to refine methods for assessing mineral scarcity and environmental impacts — the supply in some metals being expected to be a bottleneck to the energy transition —, and to identify the appropriate climate metrics for planning sustainable carbon neutrality pathways at the organizational level.

*Conclusion*: Organizations are currently facing difficulties in robustly accounting for emissions in the context of carbon neutrality goals, and these difficulties appear to be partially caused by discrepancies between standards, tools, and databases. The industrial ecology community has a key role to play in harmonizing these resources and making them more useful for planning sustainable carbon neutrality pathways.

# 1. Introduction

## 1.1. The rising popularity of carbon neutrality

The climate crisis has rapidly increased the popularity of the terms "carbon neutrality" and "net-zero", revealing serious greenhouse gas (GHG) accounting challenges for organizations — regarding methodologies, data, tools, and standards. Following the Paris Agreement, the United Nations member states have agreed to hold the increase of global mean temperature well below 2 °C and to pursue efforts to limit it to 1.5 °C (United Nations 2015). In this respect, an increasing number of public and private bodies are adopting climate targets and engaging in net-zero pathways (Net Zero Tracker 2022). As an example, more than 2000 companies already committed to setting Science-Based Targets (SBT) (Science-Based Targets 2023). Also, more than 13000 corporations, 1000 cities, and almost 100 states and regions, report disparate environmental metrics related to their transition through the Carbon Disclosure Project (CDP) (CDP 2022a). But the lack of harmonized framework and comprehensive guidance related to net-zero target-setting raise the question of the integrity of carbon neutrality claims[1] (ADEME 2022; Day et al. 2022; NewClimate Institute et al. 2022). In addition, it is unclear what challenges organizations face when aiming to quantitatively plan their carbon neutrality pathways. In parallel, research in carbon neutrality is growing rapidly. Although the term is new to many organizations, research in carbon neutrality has a long record: the term was first used in the mid-1990s by Schlamadinger et al. (1995), who paved the way for more than 900 articles (Wei et al. 2022). Despite a large academic mobilization, the industrial ecology community has mainly been involved indirectly so far by developing and discussing life cycle accounting rules (Finkbeiner and Bach 2021). Indeed, industrial ecology is an applied science whose aim is to provide sounded recommendations to decrease the environmental pressures generated by production and consumption systems. It thus relies on core environmental quantification methods, developed over several decades, that must be enlightening to sketch sustainable carbon neutrality pathways, i.e., a net-zero trajectory that does not involve unacceptable burden shifts to other environmental and socio-economic impacts.

---

[1] In this study the terms carbon neutrality and net zero are used interchangeably for a state where an organization's greenhouse gas emissions are either zero or balanced out by removals of $CO_2$ from the atmosphere.

The first objective of this paper aims to shed light on the key accounting challenges experienced by organizations to plan sustainable carbon neutrality trajectories. The second objective of this paper is to detail how this community can help address the accounting problems experienced by organizations, and support standard developers and providers. In a nutshell, this paper aims to contribute to highlighting operational challenges to high-quality accounting activities underlying decarbonization and carbon neutrality plans, and ways forward for the standard developers and industrial ecology community to alleviate or solve them. Some of these ways forward may not enjoy full acceptance in the industrial ecology community. Instead, they should be seen as recommendations emerging from the critical consideration of the authors, which call for a further level of contribution from the entire community.

## 1.2. Methodology

The methodology of this article first relies on the identification of accounting difficulties in implementing net-zero development plans reported in gray literature and by diverse organizations including some partners of the CIRAIG, a research center pioneering in sustainable life cycle metrics and systemic modeling. Public and private organizations of different sizes from around 700 to 114000 employees, and from different sectors directly reported challenges to CIRAIG through different channels. Certain challenges have been reported over time by organizations collaborating with CIRAIG, through research projects or environmental consulting projects. Other challenges have been reported during a workshop dedicated to carbon neutrality plans and organized by the CIRAIG's International Research Consortium on life cycle assessment (LCA) and Sustainable Transition. This workshop happened in November 2021 and involved 11 organizations from five economic activity sectors (manufacturing, recycling, oil & gas, electricity, information technology). These challenges will be indicated in italic.

This preliminary identification of challenges was complemented by a literature review on GHG accounting guidance and industrial ecology publications conducted in spring 2022. First, we checked whether these challenges had also been reported, and potentially elaborated, by think tanks, groups of companies such as industrial unions, or organizations developing quantification standards. To do this, we sifted through the gray literature around each challenge using the Google search engine. Specifically, we combed through the challenges and responses already provided by the extensive documentation made available on the GHG Protocol website, as the GHG Protocol is the current main standard for GHG accounting, referred to in the context of carbon neutrality by the SBTi. Second, we completed this literature review by screening the academic analyses and potential responses provided on each challenge using the ScienceDirect platform (sciencedirect.com). Finally, this last research pinpointed challenges that had not been highlighted in the context of CIRAIG's activities.

## 1.3. Carbon neutrality planning framework and challenges

Our study is structured following a framework proposed to develop sustainable carbon neutrality plans and presented in this section. A common mitigation hierarchy to address carbon neutrality relates to the generic sequence "Avoid-Reduce-Compensate", historically developed to manage environmental impacts, with some steps sometimes added or called differently (Finkbeiner and Bach 2021; Andrews 2014; Renaud-Blondeau 2022). But this sequence especially raises questions about the definition and good practices around "avoid" and "compensate" (Bach and Krinke 2021), and a clear and specific framework seems still missing for the precise development of organizational carbon neutrality plans. Moreover, this sequence does not specifically address the question of burden-shifting generated by mono-criterion environmental strategies such as carbon neutrality pathways. Any organization's sustainable net-zero strategy should then be built on the sequence "Measure-Reduce-Neutralize-Control" (MRNC), a sequence that has been implicitly used by different authors (BSI 2014; CDP 2021; Day et al. 2022). More specifically, this MRNC sequence, illustrated in Figure 1, means:

1. MEASURE: Evaluate past and current organizational GHG emissions on specific scopes.
2. REDUCE: Set net-zero objectives related to explicit timeframes and define a consistent strategy to guide mitigation actions towards a clear reduction trajectory.
3. NEUTRALIZE: Manage residual emissions, i.e., quantitatively plan and adopt solutions to neutralize emissions that have not been abated.
4. CONTROL burden-shifting: Monitor geographical and time displacement of GHG emissions, as well as other environmental, social, and economic aspects of sustainability, on the path to carbon neutrality.

This sequence is recursive: the initial "measure" step must occur regularly after a net-zero plan has been launched, to enhance the quality of the assessment with higher-quality data, measure target progress and potentially adjust the initial "reduce" and "neutralize" plans. The CONTROL step can be added to the net-zero planning process, to ensure that sustainability goes beyond the sole climate change target, and minimize the risk of burden-shifting. Each step involves specific accounting and quality challenges that have been addressed to varying extents by the industrial ecology community.

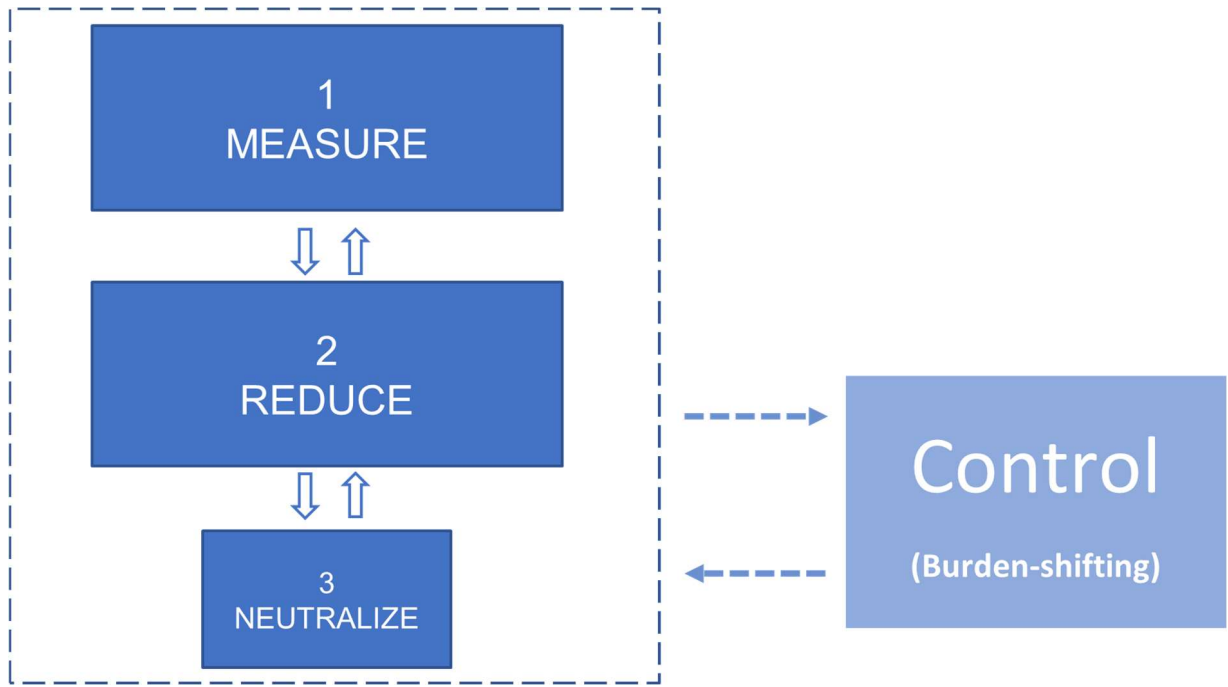

**Figure 1** Generic four-step framework to plan a (sustainable) carbon neutrality pathway

In Table 1, we give an overview of the main accounting challenges raised by organizations when planning carbon neutrality trajectories. The identified 24 challenges are structured along the four steps of the MRNC sequence and will be discussed in four sections, supported by literature references, in the light of existing standards and resources, followed by propositions of current or future solutions to these accounting problems.

**Table 1** Overview of accounting-related challenges reported by organizations, structured according to the four-step MRNC sequence presented in Figure 1

| Subsections | Challenges |
|---|---|
| **MEASURE** *(see section 2)* | |
| *Goal, scope, & other methodological choices* | #1—System boundary settings |
| | #2—End-of-Life allocation |
| *Emission inventories* | #3—Scope 1 data |
| | #4—Scope 2 quantification |
| | #5—Upstream scope 3 data |
| | #6—Downstream scope 3 modeling |
| | #7—GHG flow completeness |
| | #8—LULUC inventories |
| | #9—Uncertainty analysis |
| *Impact assessment* | #10—GWP characterization methods |
| | #11—LULUC radiative forcing |
| *Interpretation* | #12—Robustness and credibility |
| | #13—Comparability |
| **REDUCE** *(see section 3)* | |
| *Set targets* | #14—Diversity and credibility of targets |
| *Reach targets* | #15—Dynamic GWP |
| | #16—Dynamic inventories |
| | #17—Qualification as reduction actions |
| | #18—Financial planning of reduction actions |
| **NEUTRALIZE** *(see section 4)* ||
| *Internal negative emissions* | #19—Efficiency uncertainty |
| | #20—Cost uncertainty |
| *Offset's negative emissions* | #21—Quality |
| **CONTROL (burden-shifting)** *(see section 5)* ||
| *Spatial and temporal* | #22—Scope, location, sector, organization & time burden-shifting |
| *Environmental categories* | #23—Transfer towards other environmental issues |
| *Sustainability pillars* | #24—Social & economic impacts of decarbonization |

*LCI = Life Cycle Inventories; GWP = Global Warming Potential*

## 2. "Measure": GHG accounting challenges and solutions

For the "Measure" step, we make a distinction between (i) the initial measurement aiming to understand the current emissions of an organization to set the baseline to take action towards a clear reduction trajectory, and (ii) the prospective measurement necessary to project emission evolutions in the "Reduce" and "Neutralize" steps. In this section, we mainly address accounting challenges for the initial "Measure" step. Some of the identified accounting challenges are also relevant for the subsequent "Reduce" and "Neutralize" steps, as explained in Sections 3 and 4.

### 2.1. Goal, scope & other methodological choices

**System boundary settings**

CHALLENGE #1: *Organizations are frequently unsure what activities they should report on, i.e., what system boundaries to select*.

The GHG Protocol sets the reference perimeters for GHG reporting, which breaks down into scopes 1, 2, and 3, either at the corporate or product level (WBCSD and WRI 2004, 2011a). Scope 1 covers the direct emissions from the activity of the organization. Scope 2 relates to the direct emissions from an entity's purchased (and consumed) electricity, heating, steam, and cooling. Scope 3 encompasses the rest of the emissions linked to the organization's production. Upstream scope 3 emissions cover purchases embodied emissions, including indirect emissions from the production of purchased and consumed electricity (Sotos 2015a). As an example, an organization consuming electricity from Quebec's grid would report 0.6 gCO2eq/kWh under scope 2 (HydroQuébec 2021) and 33.9 gCO2eq/kWh under upstream scope 3 (HydroQuébec), which is regularly not well understood by organizations that tend to include in scope 2 the entire life cycle emissions from the electricity purchased, and this frequent confusion needs to be dispelled. Downstream scope 3 emissions are due to the transportation, usage, maintenance, and end-of-life (EoL) of the organization's production. Reporting Scope 3 is originally optional but now strongly encouraged by the GHG Protocol guidelines (resp. WBCSD and WRI 2004, p. 25, and 2011a, b, p. 6).

Some actors argue that only scope 1 should be considered in carbon neutrality plans because it is hard to control and verify scopes 2 and 3 emissions (Kaplan and Ramanna 2022), but also to avoid double counting (Lloyd et al. 2022). Indeed, scope 1 emissions of all actors can theoretically scale to global emissions if all the organizations report, as the scopes 2 and 3 of someone are the scope 1 of others (Hertwich and Wood 2018). Moreover, scope 2 and even more scope 3 are said to be harder to control by organizations so less useful for decarbonization plans. Finally, the complexity and uncertainty of scope 3 emissions can be argued to report only scopes 1 and 2 (Lloyd et al. 2022).

Nevertheless, as many organizations do not disclose yet their emissions at the product or organization levels, those that do should account for their entire life cycle to create a positive ripple effect by encouraging other companies in their supply chain to be more virtuous. This is especially right because each scope presents a substantial contribution to climate change at the global scale, as also shown at the sector level (Lloyd et al. 2022). Scope 2 is estimated to contribute between 11% (Hertwich and Wood 2018) and 25% (Victor et al. 2014) of the worldwide economy's emissions depending on the quantification method. When looking at the global economy organized in the five IPCC sectors—energy supply, transport, industry, buildings, and agriculture and forestry—scope 3 emissions encompass 52% of the worldwide economy's emissions, and the share is increasing (Hertwich and Wood 2018). Finally, scope 1 is the historical scope considered in GHG accounting and thus better understood by organizations, but only encompasses 37% of the emissions using the same IPCC sector desegregation method as discussed previously (Hertwich and Wood 2018). As each scope may generate substantial emissions, most accounting standards encourage or require an entire LCA. For instance, PAS 2060 recommends that any emission source exceeding 1% of the total emissions of the assessed system be included in the assessment scope, or that the reasons for exclusions be documented (BSI 2014). Organizational LCA (O-LCA) guidelines also claim that, ideally, the entire value chain—i.e., scopes 1, 2, and 3—should be considered within the system boundary (UNEP/SETAC 2015). The GHG Protocol as well as the SBTi standard also advocate for reporting on the entire life cycle of companies' production (WBCSD and WRI 2004; SBT 2021). Arguing the complexity and uncertainty of scope 3 is right but a weak argument face to existing tools (see section 2.2).

WAYS FORWARD #1: *Corporate emission reporting should take a life cycle perspective to generate ripple effects throughout the supply chains, and standards should align with this position. The industrial ecology community may work on operationalizing scopes 2 and 3 accounting by providing methods, tools, and data. Moreover, scope 2 might be redefined to account for the entire life cycle emissions from electricity, steam, heating, and cooling, as presented in* Figure 2*, for accounting convenience purposes, or, alternatively, clarified for stakeholders.*

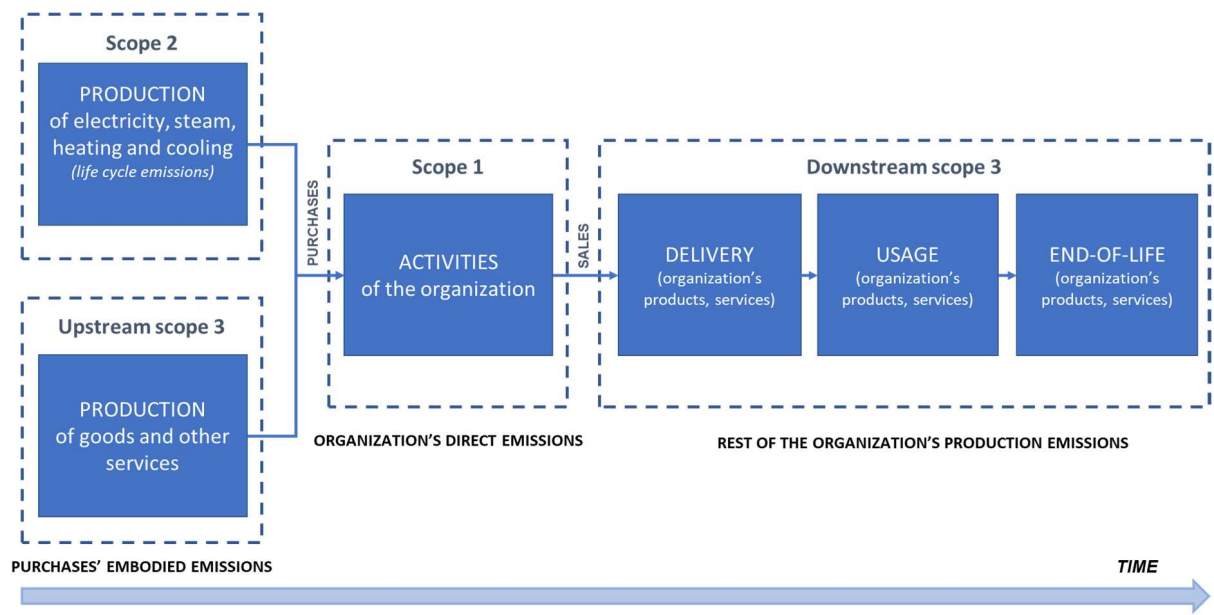

**Figure 2 Reporting scopes from the GHG Protocol, reorganized according to an organization's life cycle perspective**

**End-of-Life allocation**

CHALLENGE #2: *Guidelines providing EoL accounting rules are methodologically inconsistent.*

Organizations reported *methodological inconsistencies* between different accounting resources, such as the GHG Protocol and the LCA methodological approaches (e.g., International Organization for Standardization 2006a, b). The GHG Protocol would "*only assess the GHG emissions of the EoL before material recovery*", which could "*result in significant GHG emissions from this stage*" (WBCSD and WRI 2004, 2011b). Say otherwise, it penalizes organizations making efforts on the recyclability of their products, by not rewarding them with allocating to their products a part of the GHG avoided from recycling. In fact, for corporate accounting, the GHG Protocol includes in downstream scope 3 the emissions from the EoL treatments of sold products, and also allows considering reduced emissions from using secondary materials (instead of primary materials) and energy in the upstream scope 3 and scope 2, by not detailing the rules to account for these cases (WBCSD and WRI 2004, 2011b). However, the GHG Protocol does not allow system expansions, i.e., to allocate to the organization producing recyclable materials the potential emissions avoided from not producing equivalent virgin materials. In other words, the GHG Protocol adopted the most popular cut-off, 100:0, or "recycled content" approach to model EoL for organizations, i.e., 100% of the gains due to recycling are allocated to the consumer of the secondary materials. This method is also used in PAS 2050 for GHG accounting (BSI 2011). For products, the GHG Protocol recommends either the cut-off allocation or the recyclability substitution approach (WBCSD and WRI 2011a), also called EoL recycling approach or 0:100 approach, where the impact of the recycling process is allocated to the product itself. On the other side, LCA offers at least eleven rules to allocate losses and gains due to recycled materials between the producer and the consumer, from 0 to 100% of the impacts allocated to either the producer or the consumer, even allowing to give 100% of the impact to both (Allacker et al. 2017). Finally, allocation relies on normative choices which are still largely debated in the community and that can influence the market (Allacker et al. 2017). For example, a cut-off approach stimulates recycled material consumption, whereas the EoL recycling approach stimulates the recovery of material. Thus, rethinking these rules in the context of carbon neutrality is crucial.

WAYS FORWARD #2: *First, the consequences of EoL allocation choice must be further analyzed in the light of carbon neutrality to critically evaluate the suitability of the cut-off approach advocated by the GHG Protocol. Recycled material markets and recycling processes must be regularly assessed at local and global scales, to produce subsequent recommendations of EoL choices incentivizing allocations leading to the biggest GHG emission reductions.*

## 2.2. Emission inventories

**Scope 1 data**

CHALLENGE #3: *Organizations may have a hard time measuring on-site emissions or finding the most suitable emission factors (EFs). They report difficulties into consolidating data related to their activities.*

Scope 1 is said to be the easiest to report by organizations, but some difficulties remain in how determining EF. Background databases refer to datasets of unitary-scaled inventories of generic processes (i.e., products or activities), or by extension, to their EF. An EF is defined by the IPCC as "a coefficient that quantifies the emissions or removals of a gas per unit activity" (IPCC 2019). One part of their component remains immutable: that linked to physical and chemical rules such as the stoichiometry of the combustion of fossil fuels involved in the process. The rest depends on combustion efficiency. A geographic zone can have zero to several sources of scope 1 emission factors. Countries with cap-and-trade markets request specific EF to be used, such as Quebec where EF from LégisQuébec must be used (LégisQuébec 2021). Nevertheless, even these legal EF databases can be enhanced. For instance, LégisQuébec authorizes to complete its EF database with other sources of data, while the other sources, even when issued by governmental bodies, can be inconsistent on the same types of fuel burned.

In the case of scope 1, EFs relate to the fuels burnt on-site and other direct emissions. The difficulties to get reliable data on fuel consumption especially affect organizations with a low level of digitization, high number of sites or activities, various subsidiaries, or evolutive portfolios and production activity levels, according to what organizations reported. Organizations also highlight the potential lack of reliability of the activity data reported at the site level which would require additional verification, consolidation, and staff training to reduce mistakes. They also note the lack of consistent accounting over calendar years, as financial reporting is rarely carried out over 12 months. These difficulties also reach scopes 2 and 3 but already affect the very core scope 1.

WAYS FORWARD #3: *Primary data completeness checks and further verifications can be expected to improve with reporting quality requirements increases, as the level of reporting expertise within organizations—including accounting companies—will increase. Moreover, the rising digitalization of organizations must be an opportunity to ease high-quality data collection. On the other side, our community should get involved in reviewing scope 1 EFs to check for potential inconsistencies, and potentially propose a harmonized approach.*

**Scope 2 quantification**

CHALLENGE #4: *Two main calculation methods rule scope 2 emissions accounting, and some organizations may be confused about the methodology to adopt.*

As mentioned, scope 2 is easier to estimate than scope 3. Nevertheless, the GHG Protocol defines two different approaches for scope 2 accounting: the location- and the market-based approaches. The market-based method reflects "emissions from electricity that companies have purposefully chosen" (Sotos 2015a), through the purchasing of renewable energy certificates. On the other end, the location-based approach reflects "the average emissions intensity of grids on which energy consumption occurs (using mostly grid-average EF data)" (Sotos 2015a). According to the SBTi standard and the World Resource Institute (WRI), both location- and market-based emissions must be reported (SBT 2021), to respectively represent both "what the company physically puts in the air" versus the emissions for which the company is responsible for based on its purchases (Sotos 2015b). While the environmental impact pictured with the two approaches may indeed be more comprehensive, it doubles the quantification work and may confuse stakeholders. Additionally, it may incentivize organizations to highlight the lowest emissions in their reporting (market-based if RECs have been purchased and otherwise location-based), potentially giving an overly optimistic impression of total emissions from the provisioning of electricity to organizations.

Moreover, the market-based method generally leads to inaccurate emission reporting (Bjørn et al. 2022), which is especially problematic in the context of carbon neutrality plans. Indeed, the market-based approach can only represent the emissions an organization is responsible for when the environmental gain from the purchase of so-called green electricity is proven. Some researchers consider that this requires a consequential LCA (CLCA) (Soimakallio et al. 2011), i.e., reflecting the marginal impact of specific suppliers through certificate purchasing. But the electricity impact has been for now assessed using either attribution LCA (ALCA) or CLCA (Soimakallio et al. 2011), and there is no consensus on which approach to adopt to date (Brander et al. 2019), which results in highly uncertain impact scores (Weber et al. 2010) and confusion.

WAYS FORWARD #4: *the location-based approach must be prioritized as long as the market-based calculation rules relating to emissions from renewable electricity supply have not been clarified. But the market-based approach may complement and even replace this approach in the future if the positive effect of RECs can be robustly proven. In particular, the implications of methodological choices, such as ALCA vs. CLCA, must be investigated. Finally, deep work on energetic product LCA—and especially electricity—must be undertaken, potentially using existing standardized frameworks* (Weber et al. 2010; Heath and Mann 2012)*, to better apprehend scope 2 emissions.*

**Upstream scope 3 data**

CHALLENGE #5: *Organizations report extreme difficulties in estimating their upstream scope 3's emissions, with challenges in terms of availability and quality of the data.*

Organizations report that upstream scope 3 emissions are "*extremely challenging*" to estimate. A part of this challenge relates to the lack of primary data from suppliers. Organizations therefore rarely have access to the carbon footprint of their purchases, despite trying to involve their suppliers to get these data (CDP 2022b). Indeed, only 2% of suppliers of the CDP-reporting companies provided their life cycle footprints in 2021, and consequently, only 20% of the organizations reporting to the CDP accounted for their upstream scope 3 (CDP 2022b).

To address the lack of primary data, tools building on generic datasets related to LCA and environmentally extended input-output (EEIO) analysis have been developed. The GHG Protocol is the leading standard for upstream scope 3 estimations. It offers a free and user-friendly "scope 3 evaluator" built on Quantis SUITE 2.0 software while advising to use any other tool needed in addition to this database (Quantis 2012). This tool uses different sources of data, mostly EEIO data from the World Input-Output Database (Timmer 2012) and the US Open IO database (TSC 2011), and, to a lesser extent, ecoinvent v2.2 (Quantis 2021). Thus, the background data is mostly older than ten years, i.e., of poor quality according to the Pedigree Matrix (Weidema 1998; Ciroth 2013), and the tool does not address data quality or uncertainty. The GHG Protocol also references 55 databases that rely on EEIO, LCA, or hybrid LCA to estimate scope 3 impacts (GHG Protocol)—some of them being notably outdated. New tools addressing the scope 3 challenge have also been developed but not yet referenced by the GHG Protocol, many of them based on EEIO and offering entire life cycle accounting.

Facing this abundance of resources, organizations encounter difficulties in identifying the best sources of data, the data available often being generic and only offering a "first, essential step" or a "starting point" according to the GHG Protocol itself, to estimate upstream scope 3 emissions, before pursuing an evaluation of better quality (Quantis 2012). This is also explained by the lack of quality and uncertainty assessment in most of the resources available, and the proprietary stakes in the databases addressing them, such as ecoinvent.

WAYS FORWARD #5: *Faced with the difficulty for organizations to define the best data to assess their upstream scope 3, all the available data sources must be mapped, and their quality assessed—for example using the Pedigree Matrix approach—possibly by region and industry, to supply organizations with a harmonized good-quality open-access database and prioritize the development of missing or outdated life cycle inventories. Prioritization can be made by each sector depending on their climate change hotspots, and leading to second-generation voluntary LCI development from producers. Proprietary stakes oppose the massive and high-quality assessments necessary to* *correctly plan decarbonization, and we need to move towards shared, participatory and open-access developments, such as those offered by the Bonsai community* (Bonsai). *A first generation of datasets could be developed from EEIO using existing databases such as EXIOBASE* (Stadler et al. 2018), *USEEIO* (Yang et al. 2017), *or OpenIO-Canada* (Agez 2021) *as they are open access, and new databases should aim at better covering potentially misrepresented countries such as developing countries* (World Bank 2020). *Then, the community could aim at developing new sets of data calculated with LCA, based on a potential carbon-neutrality standardized method, to better consider intra-sectorial heterogeneities and the different efficiencies between suppliers.*

**Downstream scope 3 modeling**

CHALLENGE #6: *Many organizations are reluctant to account for downstream scope 3 because it is "very challenging", sometimes even "absolutely complex" and highly uncertain.*

Organizations report downstream scope 3 to be "*very challenging*", and "*sometimes absolutely complex to calculate*", thus resulting in rough estimates with high uncertainties and low interest in decarbonization actions. Nevertheless, as a first step, the GHG Protocol scope 3 evaluator allows for calculating the impact of packaging, distribution, and usage of the products (Quantis 2012), but not the EoL. In this tool, the use stage, which is often a major contributor to life cycle emissions—e.g., for buildings, vehicles, electronic devices, and else—is estimated based on self-declared products' lifespans and monthly consumption, within fourteen types of energy sources, and offering around 130 national and one global electricity mixes. The main limitations of this tool include that it is outdated, the background data is static and generic, and the use stage model is too simple.

Modeling the downstream scope 3 can be strenuous due to the complexity of the phenomena involved in the use stage (e.g., in pavement LCA (Santero et al. 2011), and especially the impact of road conditions on vehicle consumption (de Bortoli et al. 2022b)), the variability of users' behaviors (Shahariar et al. 2022) and product fate at the EoL, as well as other dynamic aspects that generate uncertainty. This dynamic challenge encompasses (a) the relation between the time of production and the time of value chain emissions as well as (b) the prospective uncertainty.

When it comes to (a), the time dimension is generally not correctly addressed in GHG accounting. For instance, the GHG Protocol for corporations invites to report the emissions due to one year of activity while some of these emissions may occur before or after this year, resp. due to upstream purchases and downstream usage and EoL (WBCSD and WRI 2004). The GHG Protocol for products invites to report the period of the emissions: it first mentions it about the lifespan of products and also specifies that future efficiency improvement can be considered over the lifespan (WBCSD and WRI 2011a, p. 30), and later mentions an "EoL time" relating to the EoL duration (WBCSD and WRI 2011a, p. 37), without specifically saying of to normalize reported values.

When it comes to (b), prospective uncertainty especially affects lifespans, even more for new products, despite lab tests previsions (Goulouti et al. 2020; de Bortoli 2021), potential consumptions in real conditions (Ben Dror et al. 2019), actual waste management of products, and evolution of markets and technosphere (i.e. of the environmental impacts of the economy's production). This may call for the use of prospective LCA (pLCA) which will be explained further in section 3 ("Reduce").

WAYS FORWARD #6: *The seminal GHG Protocol scope 3 evaluator accounts for downstream scope 3 emissions but should at least be updated with better quality data—in terms of geographic and temporal representativeness especially—, a more complete use stage module, and an EoL module. It will let room for future specific tools to help calculate more finely the different stages' emissions, preferably dynamically. Moreover, reporting for the year of production all the emissions of the use stage and EoL of the products or services of an organization lead to biased information on which to plan emission pathways, and our community should engage with the GHG Protocol on how to manage the time dimension and especially the time of emission in GHG reporting.*

**GHG flow completeness**

CHALLENGE #7: *There are some inconsistencies between the list of requested GHG to account for and the tools and databases available to measure GHG emissions.*

The GHG Protocol originally requested six GHGs to be included in emission inventories (WRI and WBCSD 2013): $CO_2$, $CH_4$, $N_2O$, HFCs, PFCs, and $SF_6$. When the IPCC added $NF_3$ to their GHG list in the Fourth Assessment Report (GHG Protocol 2016), the GHG Protocol also added this substance to its list (GHG Protocol 2013). Nevertheless, many environmental databases do not include this gas. Even worse, the two databases used in the GHG Protocol Scope 3 evaluator only account for $CO_2$ in the case of the WRI database, and $CO_2$, $CH_4$, and $N_2O$ in the case of the WIOD database. There are, thus, obvious inconsistencies in the different guidelines, tools, and databases.

WAYS FORWARD #7: *Consistency must be established between the GHG Protocol guidelines, tools, and databases. In particular, their referenced databases need to be updated to account for all the GHGs requested by the GHG Protocol. Characterization factors (CFs) completeness and topicality must also be checked accordingly to ensure consistent calculations, which will be addressed later in detail.*

**LULUC inventories**

CHALLENGE #8: *Because inventories on the Land Use and Land Use Change impacts of activities are scarce, these impacts are often missing from the assessment toolboxes, leading to great uncertainties in the carbon footprint of certain sectors and products.*

Land Use and Land Use Change (LULUC), i.e., the transformation and occupation of lands, have an impact on climate change through their consequences on the stocks and flows of GHG (J. Hörtenhuber et al. 2019). However, they are reported to be "*rarely considered*" in carbon accounting. This is partly due to a lack of inventory data on the impact of activities on LULUC, i.e., the lack of knowledge on the LULUC type and surface involved for a specific activity unit. It can come from a lack of guidance, resources, or motivation to collect data, or even a desire of hiding issues that irritate public opinion, particularly in case of non-popular or non-efficient practices such as deforestation-related activities (Potma Gonçalves et al. 2018) or shrimp farming (Järviö et al. 2018) - a desire that combines well with the lack of guidance on the consideration of LULUC, for instance in the current GHG Protocol documentation. Nevertheless, a new guidance addressing these issues is expected to be published in 2023. This guidance must become the new LULUC accounting standard for companies, including in programs such as the SBTi (GHG Protocol 2022). The success of this standardization will depend on both the clarity of the rules issued, their adoption in the most popular net-zero programs such as the SBTi, but also their expression in the turnkey tools offered to organizations.

In the meantime, some LULUC inventories are already available. Many initiatives—such as the organization environmental footprint sector rules guidance (European Commission 2018)—recommend following PAS 2050:2011 (BSI 2011) and supplementary PAS 2050-1:2012 (BSI 2012), and using default land-use change values in Annex C of PAS 2050, unless better data is available. It also recommends including emissions of soil carbon, excluding accumulated carbon uptakes, and reporting on the side soil carbon storage with proof of efficiency (Baitz and Bos 2020). More recently, the ISO 14067 dedicated to GHG accounting for products and organizations was published. It devotes its annex E to emissions and removal due to LULUC and also addresses the issue of inventory time profile. Other resources are likely to address this inventory issue in the future.

WAYS FORWARD #8: *A first set of solutions exist to account for LULUC from the inventory side, and they need to be investigated. The effects of activities on LULUC will need to be transparently collected, which must be facilitated with the dedicated GHG Protocol guidance to come, and its expected adoption in the most popular net-zero programs and ready-to-use accounting tools.*

**Uncertainty analysis**

CHALLENGE #9: *While a growing number of organizations understand the importance of quantifying and reporting uncertainties in environmental quantifications, there is still a global lack of uncertainty consideration in GHG reporting, and the tools proposed by the GHG Protocol or others are often not understood or/and used.*
Organizations report that "*uncertainty is rarely well accounted for in corporate reporting*", while the necessity of uncertainty in environmental quantifications to take sounded decisions is better understood. Incidentally, the GHG Protocol's Product standard requires that companies report uncertainty, at least qualitatively (WBCSD and WRI 2011a), while the corporate and scope 3 standards simply advocate for respectively managing data quality (WBCSD and WRI 2004) and perform and report qualitative uncertainty (WBCSD and WRI 2011b). Two GHG Protocol's tools and related guidance help considering uncertainty, one being general on quantitative inventory uncertainty and the other indicated to be specific to scope 3 on the GHG Protocol website, but being in fact general too but updated (GHG Protocol 2003, 2011a). These guidances present an overview of most of the mathematical techniques to quantify inventory uncertainty, and the tool follows an inventory quality scoring system derived from the Pedigree Matrix (GHG Protocol 2011a), a common approach used in LCA software and databases (Weidema and Wesnæs 1996; Ciroth 2013). But the tool proposes to allocate not defined subjective scores from very good (1) to poor quality (4) (GHG Protocol 2011b), instead of the also limited but at least quantitatively defined scores of the Pedigree Matrix. The GHG Protocol also suggests default uncertainty factors, allowing to calculate results uncertainty as well as analyze scenarios, but organizations question the source of these factors, ranging from 1 for very good data to 1.50 or 2.00 for poor quality data, and a growing number of LCA researchers disapprove the use of non-transparent or by-default uncertainty factors (Bamber et al. 2020; Heijungs 2020). In a nutshell, the GHG Protocol proposes different guidances that are partly inconsistent—between them and with the state of the art on uncertainty quantification—and organizations show reluctance to spend resources in understanding these mathematically heavy guidelines.

WAYS FORWARD #9: *Various GHG Protocol documents require or recommend uncertainty to be addressed and offer guidance and tools. But they need to be updated and the tools must be made more user-friendly, fast, and reliable, to encourage organizations to consider uncertainties.*

## 2.3. Impact Assessment

**GHG characterization methods**

CHALLENGE #10: *GWP values evolve and organizations sometimes do not use the last versions made available.*
The IPCC provides the reference GWP to consider in GHG accounting. To date, it reported six series of GWPs, from 1990 to 2021. The GHG Protocol recommends using the latest GWP, but authorizes to use previous versions (GHG Protocol 2016). In 2022, it was still acceptable to use the 2013 version, but the GHG Protocol Scope 3 evaluator (Quantis 2021) as well as some organizations in their CDP reports used earlier versions. Yet, an organization's carbon footprint can be sensitive to the underlying GWPs. For instance, the carbon footprint of activities emitting massive $CH_4$ would benefit from selecting an old series of IPCC GWP, as this gas' GWP evolved from 18.3 to 29.8 for "fossil $CH_4$ GWP-100a" (IPCC 2021). F-gases GWPs evolved even more. Using outdated GWPs distorts the climate change responsibility assessment of organizations, whose impact depends on the mix of the GHGs they release and the GWP used for each of these GHG. If GWP must keep on being used, consistency between the assessment period and the GWP series used must be ensured. At least, accounting reports must transparently specify the characterization method selected.

Nevertheless, while GWP-100a (i.e., GWP with a 100-year time horizon) are the most popular CFs, different time horizons exist (Levasseur et al. 2010) and should be rediscussed in the context of carbon-neutrality targets, building on the work of Abernethy and Jackson that advocates for a 24-year time horizon (2022). Furthermore, the Global Temperature Potential (GTP) is an alternative or complementary indicator to GWP (Levasseur et al. 2016) that can be more adapted to carbon neutrality plans as it directly relates to global mean surface temperature changes (Chang-Ke et al. 2013). Other metrics, distinguishing between the stock- and flow-like behaviors of GHGs with different atmospheric lifetimes, have since been proposed (Collins et al. 2020; Lynch et al. 2020).

A specificity involves GWP of biogenic $CO_2$, sometimes called GWPbio (European Commission. Joint Research Centre. Institute for Energy and Transport. 2014). Indeed, bio-based products such as those made of wood generate $CO_2$ emissions and removals along their life cycle, including various aspects such as forest carbon gaps and temporary carbon storage in these products (Cherubini et al. 2009). How to characterize biogenic $CO_2$ fluxes is non-consensual (Cherubini et al. 2012), but must integrate correction factors or payback times (European Commission. Joint Research Centre. Institute for Energy and Transport. 2014), including the site-specific impacts on albedo dynamics of cultures (Cherubini et al. 2012). We expect consensual recommendations to be given on some of these aspects in the GHG Protocol report dedicated to LULUC to come in 2023 (GHG Protocol 2022).

WAYS FORWARD #10: *Transparency on the characterization method used in GHG accounting reports is a first requisite, but the most recent IPCC GWPs must be quickly operationalized to allow organizations for use by organizations in the "measure" step. Moreover, further scientific discussions must address the question of metrics and time horizons to consider when assessing climate change contribution in the context of carbon neutrality pathways. In particular, shorter time horizons for GWP as well as alternative metrics such as GTP must be considered based on previous research.*

**LULUC radiative forcing**

CHALLENGE #11: *The Land Use and Land Use Change characterization factors are missing or restricted, limiting the LULUC impacts accounting.*

LULUC impacts on climate change may be missing from climate change contribution assessments due to a deficiency in life cycle inventory assessments (LCIA). The CFs for LULUC relate to GHG emissions and removal attributed to each kind of LULUC over one square meter as well as albedo impacts for biogeophysical metrics. The problems with characterization methods related to (1) the complexity of phenomena—such as the biogeophysical feedback of different forests on climate forcing (Bonan 2008)—leading to incomplete methods to calculate the impact of LULUC on climate change, or (2) even a simple failure to account for existing LULUC CFs in some climate change assessment methods like TRACI 2.1 (Bare 2011). To illustrate (1), the GHG Protocol refers to some default values while recognizing the regional specificity of the impacts (WBCSD and WRI 2011a, p. 118). Moreover, to our knowledge, the surface biogeophysics—e.g., albedo effects—is still missing in climate forcing models and would mislead climate action (Bright et al. 2015). Additionally, (2) can be illustrated by the European EN 15804 standards relating to Product Category Rules for built systems that only requested LULUC to be considered in a climate change indicator in its second version (European Standards 2019), accounting for $CO_2$, CO and $CH_4$ emitted from land transformation as well as $CO_2$ stored in soil and biomass. The inclusion of LULUC CFs has changed the environmental performance of some construction products, leading to construction-related organizations orally reporting mistrust concerning the robustness of environmental assessments. Note that just as the biophysical implications of LULUC (e.g., albedo changes) affect the climate, other indirect pathways exist, e.g., from emissions of acidifying compounds to reduced carbon uptakes in plants.

WAYS FORWARD #11: *More complete models on the impact of LULUC on climate change must be developed or implemented for GTP—especially on surface biogeophysical feedback—to better support carbon neutrality plans.*

## 2.4. Interpretation

**Robustness and credibility**

CHALLENGE #12: *Organizations report internal mistrust towards GHG accounting, lack of verification by third parties, and finally a general concern about the robustness of the results.*

Some organizations report a "*mistrust around GHG reporting inside and outside the companies that can notably affect employees' engagement and reputational aspects around climate exemplariness*". Besides the lack of uncertainty consideration discussed in challenge #9, the "*third party verification process does not challenge data quality or uncertainties*". Carbon reports are sometimes used by investors to constitute their portfolio: they would "*not do the maths*", i.e., ignore or poorly understand the complexity of $CO_2$ accounting and reporting, the uncertainty on the results, and sometimes would not even read the CDP, what would hurt the purpose of GHG reporting and CDP (Bolay et al. 2022). Let's recall that uncertainty does not only affect inventories, but occurs at each stage of the LCA, from the goal and scope definition to the interpretation of results (Huijbregts 2001).

WAYS FORWARD #12: *The tool proposed in WAYS FORWARD #8 will be a first step to account properly for uncertainties and clarify their importance. A potential standardization of GHG calculation methods and tools in the context of carbon neutrality, including reinforcing (third party) verification—would also help reduce mistrusts by enhancing GHG accounting.*

**Comparability**

CHALLENGE #13: *Reporting results are not usable to compare competitors' environmental efficiency and sectorial emission targets.*

The SBTi offers two different metrics on which to monitor GHG reductions and target reaching: (1) economic intensity metrics, considering the emissions per functional unit (i.e., production unit, that can be physical or economic) and (2) absolute metrics, quantifying the emissions of the whole company/activity (SBT 2019). In the first case, companies calculate their carbon intensities per production unit, and stakeholders tend to consider these intensity metrics as a climate performance indicator and compare these intensities between companies of a common sector.

Yet, accounting quality is overall very low, when cumulating heterogeneity of guidelines interpretations and thus accounting methods, as well as poor data and CF quality. Thus, organizations report accounting as "*useful for identifying hotspots and finding areas of improvement*" but "*not useful for comparing with other companies or measuring progress over time*". Indeed, carbon footprints are not comparable between organizations. Thus, using competitors' environmental efficiency to differentiate them is questionable, even though the only current way to consider impacts on climate. But the potential "optimization" of carbon footprint performed by organizations by choosing beneficial accounting rules to attract customers, raise investments, or decrease potential carbon fees in places with cap-and-trade systems must be stopped.

WAYS FORWARD #13: *Standardization of calculation methods, tools, and databases with strong verifications seems the only way to obtain reliable climate performance indicators from reporting and make emission metrics and targets meaningful, but also comparable within a common sector for stakeholders.*

## 3. "Reduce"

The "reduce" stage of the MRNC sequence includes first to set targets — i.e., choosing emission reduction targets — and second to develop an adequate plan to reach these targets. The challenges encountered in these two aspects of the stage are different, and detailed in this section.

### 3.1 Set targets

CHALLENGE #14: *Some organizations communicate non-accredited reduction targets or have difficulties choosing their targets, among the diversity of programs and indicators, resulting in non-comparable, non-* *credible, and/or non-ambitious targets.*

Organizations can freely set their carbon-neutrality targets, i.e., the temporality and importance of the reductions targeted. This makes assessing the real contribution of the organization's transformation efforts associated with carbon neutrality difficult. To increase the impact and credibility of their plan towards carbon neutrality, organizations can join the race to zero campaign, a United Nations Climate Change initiative aiming at accelerating decarbonization. This initiative identifies programs to commit to credible reduction targets towards carbon neutrality for all types of organizations (UNFCCC). Each kind of organization can refer to different programs—also called "accredited partners", and some may refer to more than one program. It is the case for companies (five programs), cities (two), and universities (two) (UNFCCC). In 2022, companies could choose between Business Ambition for 1.5 C—Our Only Future (SBT, 1500 companies enrolled), Business Declares (Business Declares, 87 companies), The Climate Pledge (The Climate Pledge, 315 companies), Exponential Roadmap Initiative (Exponential roadmap, 50+ companies) and Planet Mark (PlanetMark, 389 companies). Some organizations reported the Exponential Roadmap Initiative's and the Climate Pledge's targets as "not sufficiently well defined", contrary to the SBTi business ambition for 1.5° that proposes targets for several sectors (SBT 2021). The SBTi also offers a tool to calculate medium-term targets towards carbon neutrality, proposing different methodologies (absolute or intensity targets), system boundaries (from scope 1 only to the entire life cycle), underlying emission allocation principles, base and target years, and sectorial pathways (Bjørn et al. 2021; SBTi 2022). But any kind of reduction targets—including non-accredited ones—can be selected by companies, thus lacking comparability due to methodological variability. For instance, an intensity decrease is compatible with an absolute increase. Finally, the ambition and the genuineness of targets are globally not well understood, but some studies start exposing dubious practices, such as companies who reached their targets before implementing their reduction program (Bolay et al. 2022) or within less than two years (i.e., before the SBTi had the time to approve the targets) (Giesekam et al. 2021), demonstrating at least a lack of reduction ambition, at worst dishonest targets.

*WAYS FORWARD #14: The industrial ecology community may help identify and promote the best fitting program for the organization's context or alternative ways to set reduction targets. In particular, the scope, timeframe, and reduction indicators must be analyzed in the light of the expertise of the community to ensure target credibility, and support efficient green investments and the necessary transformation of the economy.*

## 3.2. Reach targets

The major difficulty in the "Reduce" stage may not be to set reduction targets, but rather find a way to reach them. This requires first robust quantifications of reduction alternatives, involving dynamic considerations. Indeed, the efficiency of a reduction action is generally modeled in the past, at best for the present, and rarely in the future. Six dimensions of dynamic LCA have been identified (Lueddeckens et al. 2020), GWPs and GHG life cycle inventories (LCI) being two main dimensions to robustly assess and hierarchize future actions to support transition strategies, but rarely considered prospectively. Moreover, organizations question the cost of reduction strategies as well as the tools to do so.

**Dynamic global warming potentials**

CHALLENGE #15: *Static GWPs are considered to assess future emissions, whereas they evolve, generating uncertainty in the assessment of reduction alternatives.*

The equation to calculate GWPs considers marginal mechanisms based on the average GHG concentration in the atmosphere (Reisinger et al. 2011). Thus, GWPs evolve as the GHG concentration changes. Dynamic GWPs—e.g., time-dependent GWPs—were used in a pioneering LCA by Levasseur et al. (2010) and showed to have substantial impacts on the final results. To be consistent, GWPs in the "Reduce" and "Neutralize" steps should ideally be calculated under the representative concentration pathway (RCP) considered in each carbon neutrality plan, for instance using Reisinger et al. results (2011), and uncertainty should also be considered on GWPs when assessing future actions' consequences. In a similar vein, some have proposed that the time horizon of the GWP should be defined by a fixed end year, derived from a global climate goal, and thereby decrease with time (instead of a fixed time horizon of, say, 100 years) (Abernethy and Jackson 2022). Nevertheless, such dynamic approaches would require clear guidance and adequate databases or tools to support organizations and may require additional efforts to calculate the probability of future reduction action and to support decision-making on these results.

WAYS FORWARD #15: *Some dynamic GWPs exist and may be implemented in the "reduce" step of carbon neutrality plans, but this would require that the importance of dynamic considerations be examined further to evaluate how this additional layer of complexity needs to be brought into carbon footprinting.*

**Dynamic inventories**

CHALLENGE #16: *Inventories and EFs are most often not representative of the future and organizations wonder how to manage this source of uncertainty.*

In carbon footprinting, LCIs reflect the quantities of GHGs emitted and potentially absorbed/removed by a process (=an activity). Removals are GHG flows removed from the atmosphere through man-made actions (e.g., by afforestation/reforestation or "technological" means). The carbon footprint of a process varies over time in the fixed geographic zone, as it depends on various parameters such as potentially technologies, behaviors, and background technosphere. For example, EFs of electricity mixes in different states of the US evolved between 2001 and 2017: they generally decreased but increased in some places, like Idaho (Schivley et al. 2018). Assessment of future emissions or removals should be representative of changes in the background system, i.e., the assessment of reduction and neutralization actions as well as potentially downstream scope 3. Dynamic LCIs must also be used to retrospectively account for GHG emissions and removals and validate trajectories, which is in reality never done.

So far, most academic studies published on dynamic LCA exclusively focused on dynamic electricity LCIs (Cornago et al. 2022). Considering dynamic electricity LCIs is not enough as electricity accounts for less than one-third of the worldwide emissions (2014). Yet, LCA-based tools used for scope 3 assessments are not even including that time dependency: they are completely static. Recently, some tools were developed to account for foreground dynamic inventories and generate time-dependent emissions visualization, pointing towards near-real-time LCA monitoring (Ferrari et al. 2021; Rovelli et al. 2022). Though, its dynamics only cover foreground inventories. With the intensification of research on pLCA (Thonemann et al. 2020), methodologies improved recently by using integrated assessment models (IAMs) and shared socioeconomic pathways (SSPs) (Absar et al. 2021; Pedneault et al. 2021; Kaddoura et al. 2022), that allow scenarizing optimal futures based on a GHG constraint as well as physics and economics (Wilson et al. 2021). Recently, the PREMISE tool generalized the development of complete prospective life cycle inventory (pLCI) databases, based on previous recommendations of the community to enhance the assessment of future technologies and actions (van der Giesen et al. 2020). PREMISE is a pioneering promising tool to account for dynamic background inventories. It was applied to ecoinvent v3.8 using scenarios generated by the REMIND IAM (Leimbach et al. 2010; Aboumahboub et al. 2020; Sacchi et al. 2022), and led to open-access prospective versions of ecoinvent 3.8, cut-off, based on different SSPs (Sacchi et al. 2022). Finally, the superstructure approach proposed by Steubing and de Koning (2021) to simplify pLCA calculations—which can lead to dozens of prospective scenarios—is also recent progress to help develop the necessary pLCI databases needed for carbon neutrality planning.

WAYS FORWARD #16: *Future emissions can already be estimated by using PREMISE and the superstructure approach, but no databases related to the past have been developed so far. The LCA community could keep on developing pLCA by investigating different IAMs quality and completeness in the context of carbon neutrality. Moreover, coupling EEIO with IAM is the next step to completing specific pLCA with wide-scale prospective EEIO assessments. Finally, dynamic inventories must also be developed for retrospective carbon footprinting, to validate past emissions and trajectories, both in terms of emissions and removals.*

**Qualification as a reduction action**

CHALLENGE #17: *What can be accounted for in the "reduce" step to reach net-zero targets is unclear to organizations, especially when it comes to "offset", "carbon credit", or "avoided emissions".*

Many organizations reported counting on GHG offsets to reach their reduction targets, while main guidelines disagree on allowing or not offsets in the "reduce" step. Offsets are "GHG emissions reductions made outside of the subject carbon footprint" (BSI 2014), i.e., carbon credits or avoided emissions, the latter of which are also referred to as "comparative" emissions and defined as emissions saved compared to reference products or services (Russell 2019).

First, PAS 2060 differentiates the reduction of GHG emissions from the following "offset" step that consists in managing residual GHG emissions (BSI 2014). For PAS 2060, offset should not be included in the reduction step, but can be used to neutralize residual emissions. More recently, the SBTi corporate net-zero standard excluded carbon credit and avoided emissions from admissible reduction actions (SBTi 2021, pp. 21, 42). But it accepts renewable electricity targets and consideration of "Renewable Energy Certificate" (REC)—also called renewable energy credits—and virtual power purchase agreement in the "measure" and "reduce" steps (SBTi 2021, pp. 25–28, p44), as it refers companies to the GHG Protocol scope 2 standard (Sotos 2015a; SBT 2021, pp. 25–28). Considering the inability of these certificates, so far, to truly reduce emissions (Bjørn et al. 2022), and the urgency to stabilize climate change (IPCC 2022; SBT 2021), this can be seen as inconsistent. Thus, we would not recommend including credits—including REC in their current situation of overall non-additionality—and avoided emissions in the reduce step.

WAYS FORWARD #17: *Contradictory guidance exists, and we recommend the GHG Protocol to align with PAS 2060 and the SBTi standard in banning carbon credits as a GHG reduction action. Moreover, SBTi must justify or change its position on RECs, given that the initiative does not allow purchases of the conceptually similar carbon credits to count as emission reductions.*

**Financial planning of reduction actions**

CHALLENGE #18: *Organizations report struggling in prioritizing and planning reduction actions required to meet their targets.*

Reducing GHG emissions can represent a cost that especially concerns companies, and they report having trouble to assess the future cost of reduction actions and planning them accordingly to reach their emission targets while consolidating their business plan. Indeed, investment planning is crucial for private companies that must be profitable (Moshrefi et al. 2020; Ayoub et al. 2020), and the consequences of public policies and especially carbon markets are now recognized as a real financial risk (Ayoub et al. 2020).

Corporate financial strategy planning often relies on cost-benefit analyses that are dynamic calculations highlighting more remunerative decisions. On the other hand, marginal abatement cost (MAC) curves are common indicators used to calculate the cost efficiency of emission abatement strategies (Kesicki and Strachan 2011). The main caveats of the method are the lack of uncertainty and dynamic considerations (Kesicki and Strachan 2011). MAC curves have already been calculated by hybridizing LCA and (dynamic) life cycle cost analysis (LCCA) (Butt et al. 2020; de Bortoli et al. 2022c), but not based on pLCA and dynamic assessments.

WAYS FORWARD #18: *MAC curves can be used as a tool to select the most cost-efficient decarbonization actions, by combining pLCA and dynamic LCCA, preferably considering uncertainty.*

## 4. "Neutralize"

Neutralize is the last step to planning carbon neutrality pathways, before (optional) burden-shifting control. This step aims to balance residual emissions, i.e., emissions that were not abated, and rely on both internal negative emissions—i.e., occurring within the organization's supply chain—and credits' negative emissions, taking place outside the supply chain. Terminologies on neutralization solutions are still unstable, but we can define as "nature-based" solutions those relying on the capacity of natural compartments such as soil and plants to capture and store carbon dioxide. Then, "technology-based" solutions are carbon capture and sequestration (CCS) technologies, that capture and transform emissions by means of industrial processes (Ma et al. 2022). They include bioenergy with carbon capture and storage (BECCS), GHG captured at industrial production sources, and direct air capture (DAC). DAC is marginal, as it only represented 0.025% of the 40 Mt $CO_2$eq technology-based carbon capture in 2020 (Ma et al. 2022). Note that neutralization actions may also target, over the organization's or system's life cycle, greater removals than emissions, out of ambition to go "beyond net-zero" or need to compensate for organizations with less leeway in their carbon footprint, to finally reach a carbon-neutral society.

## 4.1. Internal negative emissions

**Efficiency uncertainty**

CHALLENGE #19: *Organizations report many challenges in planning residual emission neutralization due to uncertainty in carbon-abatement efficiency.*

Organizations report difficulties in planning residual emission neutralization, due to many uncertainties, including efficiency (Markusson et al. 2011; Gabrielli et al. 2020; Ma et al. 2022). The efficiency uncertainty relates to scalability, quality of carbon capture performance assessments, and cost. The neutralization potential of these solutions must be estimated on their complete life cycle and not on the sole use stage: a complete life cycle approach, while obvious to the LCA community, is still not a standard approach within the industry. Moreover, some LCA assess the GHG neutralized by CCS (Sathre et al. 2011; Singh et al. 2011), but with limited dynamic considerations. Again, pLCA must be rallied to properly quantify the efficiency range of early stage neutralizing solutions, with suitable IAMs considering penetration rate scenarios, and uncertainty must also be considered.

WAYS FORWARD #19: *The LCA community must develop and use pLCA coupled with suitable IAMs to properly quantify the efficiency range of early stage neutralizing solutions.*

**Cost uncertainty**

CHALLENGE #20: *The economic and financial viability of neutralizing solutions is a concern for organizations that would like to better assess it.*

Organizations report difficult to look at carbon capture, and they consider it as an investment decision risk. Indeed, costs have been said to be the most significant brake to CCS deployment (Budinis et al. 2018). To better master this risk, a set of financial and economic models (Markusson et al. 2011) can be coupled to pLCA and IAMs to generate MAC curves as a decision-supporting tool. Let's note that IAMs have already demonstrated their ability to assess the cost of CCS (Budinis et al. 2018). Policy uncertainty also creates a risk to the viability of CCS and may be assessed using scenario analyses (Markusson et al. 2011).

WAYS FORWARD #20: *IAM can help understand the future cost of CCS, and can be coupled with pLCA, MAC curves, and scenario analyses to better prepare for potential futures.*

## 4.2. Offset's negative emissions

CHALLENGE #21: *Some organizations express concerns about the quality of the carbon credits they can buy, and many questions arise around their genuineness and the verification processes.*

Carbon offsetting mechanisms are multiplying rapidly. To date, more than 40 countries and 20 cities, accounting for 22% of the worldwide GHG emissions, use carbon pricing mechanisms such as cap-and-trade systems (World Bank 2020), and many companies buy REC or other carbon credits to decrease their reported GHG emissions (Kollmuss et al. 2015). Despite this rising success, the genuineness of carbon credits, e.g., their real ability to decrease emissions, has been vigorously criticized (Schneider 2009; Alexeew et al. 2010; Kollmuss et al. 2015; Cames et al. 2016; Marino et al. 2019; Seymour 2020; West et al. 2020; Badgley et al. 2021; Marino and Bautista 2022; Bjørn et al. 2022). PAS 2060 specifies the following criteria checklist to ensure the genuineness of carbon offsets: additionality, permanence, leakage, and double counting (BSI 2014, p. 26). Both PAS 2060 and the World Bank ask for the quantification methodologies to be provided and if possible to use recognized measurement tools (BSI 2014; World Bank 2020), as credits' calculated efficiency largely depends on the assessment method used (Haya et al. 2020; Finkbeiner and Bach 2021). In response to these inconsistencies, some governments such as California established standardized assessment protocols, but these protocols still led to punctual climate-negative actions due to protocol-level quality issues (Haya et al. 2020).

Moreover, the future availability of high-quality carbon credits on the market is linked to the financial risk of neutralization, mentioned in challenge #20 about internal negative emissions, but that also affects offsetting. Indeed, if the quality criteria for carbon credit matter, there is a risk that the demand for high-quality credits will exceed the offer, with the multiplication of GHG reduction requirements. This would have the effect of increasing the cost of these credits. Without a major transformation of the current energy and economic systems, supply may not be able to match demand due to the physical limit of quality offset projects, for instance in terms of space. It should also be noted that the use of credits in GHG accounting reports may spread the false idea that goods or activities can be "0 emissions", which is why the SBTi calls for significant reduction targets and accounting for offsets apart from reduction actions.

WAYS FORWARD #21: *To ensure the quality of carbon credits, a sort of standardization of credit accounting methods seems ineluctable, but these methods must be tailored for different types of offset projects to avoid negative effects, and the LCA community should participate in the standards development.*

## 5. "Control" (burden-shifting)

Carbon neutrality plans are mono-objective: they aim to support the global goal of zero or negative GHG emissions. The required emissions reduction and neutralization actions may generate impact shifts, either between locations, environmental impact categories, or towards the social and economic pillars of sustainability, which can be problematic.

### 5.1. Spatial and temporal

CHALLENGE #22: S*ome organizations are concerned about emission transfers in time and space linked to the implementation of carbon neutrality plans.*

Such emission transfers can occur between scopes, geographic regions, periods, and/or even economic sectors and companies. For instance, decarbonization actions conducted on a restricted scope rather than on the entire life cycle of the system, can lead to overall increase the entire life cycle GHG emissions. As an example, the common exclusion of scope 3 emissions in reporting could lead to bigger increases in scope 3 than the reduction occurring in scopes 1 and 2 when implementing decarbonization actions, as showcased in the infrastructure sector on downstream scope 3 (Jackson and Brander 2019). Impact transfer has also been demonstrated along the value chain or between regions by Cadarso et al. using multiregional input-output (MRIO) analysis, but traceability and modeling of imports must be enhanced to increase the representativeness of their environmental quantification (2018). Weinzettel et al. also showed that, when considering the consumption of internationally traded products, higher-income countries tend to displace a larger fraction of land use abroad, and thus pressure on foreign ecosystems in low-income countries (Weinzettel et al. 2013), which calls again for inclusion of scope 3 not to displace impacts between countries.

WAYS FORWARD #22: *Full LCA, i.e., complete inclusions of scopes 1, 2, and 3 recommended in section 2.1., prevent stealth burden-shifting in time and space. Moreover, regionalization in LCA can allow locating where emissions take place. At large scales, MRIO conveniently captures impacts all along the supply chain, nevertheless tracking of internationally multi-traded products must be enhanced to increase the quality of consumption-based quantifications.*

## 5.2. Between environmental categories

CHALLENGE #23: *Organizations question the trade-offs between GHG emissions mitigation and other environmental impact categories and how to account for them.*

Several recent papers questioned if climate and other environmental protections were conflicting goals, calling for more multicriteria assessments (Yang et al. 2012). The main burden-shifting identified so far has been shown to put a strain on metal depletion, LULUC, sometimes on human health and ecosystems in general, and more rarely on other impact categories such as water scarcity or eutrophication (Laurent et al. 2012).

First, switching from fossil fuels to cleaner energies is key to reducing GHG emissions (Yuan et al. 2022), but will require negotiating a coordinated transformation of the interlinked energy and metal systems (Wang et al. 2022a). In response, a growing corpus of studies has looked at metal resource bottlenecks in the energy transition. Depending on the perimeters assessed, the pathways, and the models considered, different metals are expected to potentially deplete, and among the recurringly distinguished metals are cobalt, nickel, lithium, silver, some rare-earth elements, and platinum-group metals (Watari et al. 2018, 2020). For instance, Junne et al. warned of the criticality of the reserves in lithium and cobalt, and potentially also dysprosium, using Material Flow Analysis (MFA) (Junne et al. 2020b), while Tong et al. expected strong pressure on platinum-group metals due to the development of fuel cells, by coupling MFA and IAMs (Tong et al. 2022). On the other hand, the rise in metal consumption will not only deplete resources but also change their impact intensity, and increase other environmental pressures. Between 1995 and 2015, metals already accounted for around 40% of the GHG released by material production worldwide (Hertwich 2021). Later, Van der Meide et al. showed by coupling LCA and the IMAGE IAM that the carbon footprint of cobalt production would increase by 9% under a business-as-usual scenario but decrease by 28% under a "sustainable development" scenario (van der Meide et al. 2022). These researchers also showed a high increase in human toxicity impacts from cobalt production (71–112%). On the other hand, Naegler et al. showed by coupling LCA and bottom-up energy system models that climate strategies were only expected to increase mineral resources consumption, land use, and some impacts on human health (Naegler et al. 2022). We can conclude from these studies that considering dynamic LCIs and prospective scenarios will be as key in multicriteria assessments as for the initial mono-criterion "reduce" step but will lead to ranges of potential consequences.

Moreover, metal depletion in carbon neutrality pathways has mainly been studied using MFA because of the limited quality of resource depletion indicators in LCA (Verones et al. 2017; Berger et al. 2020) and metal LCIs in databases. For instance, Exiobase accounts for metal consumption, but only disaggregates copper, aluminum, and steel, while most of the expected critical metals for the energy transition are aggregated in an "other metals" category (Stadler et al. 2018). Thus, progress in LCA and EEIO modeling will be needed to analyze the environmental impact of the growing metal consumption.

LULUC is the second impact to stand out after metal-related impacts. Problematic land consumption can arise when replacing fossil fuel with biomass (Junne et al. 2020a) and with relying on BECCS for negative emissions (Seymour 2020). Incidentally, researchers call for more attention to the geographic context of BECCS, and more transparent assessments of these solutions (Tanzer et al. 2021). Moreover, CDP proposes to report deforestation, but tracking impacts on the upstream supply chain is particularly difficult as only 47% of suppliers are involved in this kind of reporting (CDP 2022b).

Other environmental categories could be affected by the energy transition. For instance, the literature suggests that ethanol would present higher acidification and eutrophication impacts than gasoline (Baral 2012), while several countries such as Spain, Egypt, or China would face water scarcity as the main barrier to energy sector transformation (Berger et al. 2015; Zhang et al. 2021). Certain human health indicators such as carcinogenic and non-carcinogenic effects and respiratory effects could also deteriorate while implementing the energy transition (Naegler et al. 2022). Moreover, digitalization can help decarbonize some sectors such as the road industry (de Bortoli et al. 2023) but must also come with environmental burden-shifting: a study for ADEME identified metal resource depletion but also freshwater eutrophication as digitalization's most problematic impact transfers (Bio Intelligence Service 2011).

WAYS FORWARD #23: *LCA and EEIO can be used to assess environmental category burden-shifting, and a dynamic approach is again recommended. The community needs to enhance metal-related inventories and related* impact assessment methods but can for now complete LCA or EEIO methods by MFA to account for metal depletion at stake in energy transition and digitalization.

### 5.3. Between the three sustainability pillars

CHALLENGE #24: *Some organizations solely focus on reducing GHG emissions, and this may put at risk holistic sustainable development goals.*

A carbon-neutral plan does not equal a sustainability plan, as the latter must also encompass other environmental dimensions as well as social and economic metrics, to cover the complete triple bottom line of sustainability. Thus, solely focusing on GHG reduction risks increasing other impacts.

First, energy transition is core for carbon neutrality but generates burden shifting (De La Peña et al. 2022) that can be hard to understand and quantify. Positive as well as negative trade-offs can occur at different scales (Mancini and Sala 2018), pushing towards large-scale as well as local assessments when looking at socioeconomic burden-shifting. For instance, Hottenroth et al. quantified the impact of different decarbonizing energy system pathways on sustainability metrics — macroeconomic indicators such as gross domestic product, unemployment rate, number of new and lost jobs, as well as on social metrics such as diversity or regional inequality — finding no correlation between the reduction ambition and non-GHG sustainability metrics (Hottenroth et al. 2022). Another example is the use of biomass energy and the development of BECCS, which commonly take part in decarbonization solutions, while IPCC draws attention to the risk they could pose to food security in certain pathways (IPCC 2020). IPCC even encourages decreasing the use of traditional biomass for energy, to secure health and socioeconomic conditions of some vulnerable communities (IPCC 2020). The energy transition is also expected to be highly metal-consuming, but Watari and colleagues argue that the social and environmental implications induced by the growth in metal demand have still been too scarcely quantified (Watari et al. 2020). Mining has been shown to bring modest socioeconomic returns for local communities (Aguirre Unceta 2021), or even aggravate income and health inequalities within communities (Shiquan et al. 2022). Specifically, artisanal cobalt mining carries ethical problems as the high exposure to the trace metals in mine dust generates long-term morbidity in vulnerable communities (Banza Lubaba Nkulu et al. 2018). In terms of methods, Mancini and Sala identified four main frameworks to quantify the sustainability performance of the mining sector: the UN Sustainability Development Goals (SDG), the Global Reporting Initiative (GRI), the European Union regulation and policy impact assessment, and social LCA (S-LCA) (2018). S-LCA has been used to quantify the social or socio-economic impacts of products closely linked to many decarbonization pathways, such as critical metals, digitalization objects, and renewable energy devices (Subramanian and Yung 2018; Buchmayr et al. 2022). Schlör et al. also proposed an original method mixing social footprint, societal life cycle costing (sLCC), and the social risk intensity based on Human Development Index to quantify the impact of metal consumption in renewable energy technologies (Schlör et al. 2018). The DESIRES framework goes further by integrating an optimization layer on top of LCA, LCCA, and S-LCA (Azapagic et al. 2016). Then, monetizing externalities of mining based on LCA, such as done by Arendt et al. on environmental externalities (Arendt et al. 2022), may lead to interesting indicators to drive a sustainable net-zero future.

Incidentally, the sustainability of carbon neutrality plans also concerns specifically who will support the transition costs. For instance, decarbonized electricity generation systems generally present a higher capital expenditure than fossil-based systems (Hertwich et al. 2015), which can be a brake for organizations with short- or mid-term financial management. Obviously, the financial burden of decarbonized technologies can also reach consumers, as said by Dirnaichner et al. about hydrogen-based technologies (Dirnaichner et al. 2022). More largely, the green energy competition could reshape geopolitical connections (Yuan et al. 2022), which could strongly affect supply chains and sustainability performance.

The net-zero transition also brings the question of social acceptance. Various changes such as within the sharing economy are well accepted (Cherry et al. 2018), although this economy does not necessarily keep its decarbonization promises (de Bortoli and Christoforou 2020; de Bortoli 2021). But the future development of some renewable energy infrastructure raises concerns in terms of fairness and trustworthiness or reliability and can lead to social unrest, strikes, or NIMBY effects (Joly and De Jaeger 2021). A few examples of these locally unpopular infrastructures are mines (e.g., França Pimenta et al. 2021), dams (e.g., Piróg et al. 2019), railways (e.g., He et al. 2016), and onshore wind turbines (e.g., Peri et al. 2020). The noise of some of these infrastructures is rarely assessed but can be a major concern for residents (Peri et al. 2020), while potentially less substantially impacting than traffic noise (Radun et al. 2022). Incidentally, noise started to be recently integrated into environmental LCA (eLCA) (Meyer et al. 2017; de Bortoli et al. 2022b), and major avenues could be explored in the context of the sustainability assessment of carbon-neutral pathways, both on LCI and CFs. Finally, S-LCA and eLCA can increase transparency on socioeconomic and environmental impacts for the different stakeholders and can be coupled as performed in some studies (Chang et al. 2015; Wang et al. 2022b), but researchers also emphasize the need to create effective and meaningful social involvement (Heras-Saizarbitoria et al. 2013) and develop project visualization tools (Cranmer et al. 2020) to enhance social acceptance. More generally, the holistic sustainability of any decarbonization action to achieve carbon neutrality could be assessed using the life cycle sustainability assessment (LCSA) framework developed by UNEP/SETAC and mixing LCAs, LCAs and LCAs (Valdivia et al. 2013).

To step back, and from an individual point of view, a growing corpus of studies looks at demand, lifestyle, and happiness. It questions if 100% technology-oriented decarbonization pathways could lead to carbon neutrality due to rebound effects (Hertwich 2008) and ever-growing demand (Creutzig et al. 2018), amplified by the social promotion of unsustainable lifestyles (Weinzettel et al. 2013; Creutzig et al. 2018), while intentional simplicity could provide better life satisfaction (Vita et al. 2020).

WAYS FORWARD #24: *Several methods can be hybridized to ensure more holistic sustainable carbon neutrality trajectories, such as the classic S-LCA, LCCA, and eLCA, for instance following to the UNEP/SETAC LCSA framework. Noise methods in LCA may be an avenue to develop further around renewable energy systems. Despite their intrinsic subjectivity, sustainable indicator weighting methods, such as monetization, can be applied to help decision-makers, potentially using optimization tools. But sustainability also relates to geopolitics, and lifestyle global consequences, that must be investigated further with the help of IAMs.*

## 6. Discussion and recommendation

### 6.1. Synthesis of recommendations

Table 2 synthesizes the recommendation we made for each of the 24 accounting challenges found in the development of carbon neutrality plans. These recommendations have brought to light certain key and cross-cutting issues that we want to discuss further in this section: the consideration of time in the assessment of sustainability, the selection of attributional or consequential approaches, and the standardization of quantification methods in the context of carbon neutrality.

**Table 2 Synthesis of our recommendations answering each of the 24 accounting challenges in carbon neutrality planning**

| Challenges | Ways forward |
|---|---|

|     |                         | **MEASURE** *(see section 2)* |
| --- | ----------------------- | ----------------------------- |
| #1  | System boundary         | Redefine scope 2 & develop tools to ease robust scope 2 and 3' assessments, to consider the entire life cycle |
| #2  | EoL allocation          | Analyze the consequences of EoL allocation choice regionally to push towards the best practices |
| #3  | Scope 1 data            | Audit & enhance scope 1 EFs, enhance field data collection with digitalization & expertise |
| #4  | Scope 2 quantification  | Prioritize location-based while consolidating market-based, question ALCA/CLCA, develop electricity LCI |
| #5  | Upstream scope 3 data   | Map databases & audit quality, develop high-quality open-access databases (EEIO then LCA) |
| #6  | Downstream scope 3      | Update the GHG Protocol scope 3 evaluator, develop finer calculators, question time dimension inclusion |
| #7  | GHG completeness        | Establish consistency within the GHG Protocol & relating resources (standard, guidance, databases, tools) |
| #8  | LULUC (inventories)     | Investigate LULUC effects on climate change, and activity inventories to account for these effects |
| #9  | Uncertainty calculation | Update GHG Protocol resources on uncertainty and make the tools more user-friendly, fast & reliable |
| #10 | GWP CF                  | Operationalize the latest IPCC GWP, question time horizon, and consider alternative indicators (GTP) |
| #11 | LULUC (CF)              | Complete LULUC LCIA for climate change contribution, including biogeophysical feedbacks in GTP |
| #12 | Robustness & credibility| Reinforce third-party verification, use reference accounting standards and tools |
| #13 | Comparability           | Discuss standardization of calculation methods, tools, and databases, reinforce verifications |
|     |                         | **REDUCE** *(see section 3)* |
| #14 | Diversity & credibility | Identify and promote the best reduction target program, analyze scope, timeframe, and indicators |
| #15 | Dynamic GWP             | Consider implementing existing dynamic GWP |
| #16 | Dynamic inventories     | Use PREMISE, develop pLCA, investigate IAMs, couple EEIO & IAM, develop retrospective LCI |
| #17 | Qualified actions       | Consider banning offset from reduction actions in all standards, including REC if additionality is not proven |
| #18 | #Financial planning     | Use MAC curves by combining pLCA and dynamic LCCA, consider uncertainty |
|     |                         | **NEUTRALIZE** *(see section 4)* |
| #19 | Efficiency uncertainty  | Assess the neutralizing range of early-stage solutions with pLCA and suitable IAM |
| #20 | Cost uncertainty        | Couple MAC curves, pLCA, and IAM |
| #21 | Quality                 | Consider standardizing credit accounting methods, tailored to different types of offset projects |
|     |                         | **CONTROL (burden-shifting)** *(see section 5)* |
| #22 | GHG displacement        | Consider regionalized full LCA, enhance traceability and modeling of multi-traded products in MRIO |
| #23 | Environmental issues    | Complete LCA and EEIO with MFA, enhance and implement metal-related inventories and LCIA |
| #24 | Social & economic       | Couple S-LCA, LCCA, and eLCA, develop noise in LCA, investigate geopolitics & lifestyle with IAM |

## 6.2. Including the time dimension is necessary

The GHG corporate Protocol recommends using "consistent methodologies to allow for meaningful comparisons of emissions over time" (WBCSD and WRI 2004). But to our point of view, it partly fails in supporting such practices as it provides transparent but not updated tools, using various and sometimes obsolete calculation methods that for instance hinder the comparability of times series between companies using different methods. More globally, the time dimension related to climate change contributions is missing from most of the operational guidelines related to GHG accounting and decarbonization plans. Yet, temporal issues are acknowledged to be a major source of inaccuracy in LCA (Lueddeckens et al. 2020), and performing static GHG quantifications must lead to miscalculations and potentially biased decision-making. Time-dependency goes beyond characterization factors and LCI (Lueddeckens et al. 2020). But, in our opinion, these two aspects must be addressed as a priority to inform and plan carbon neutrality trajectories, while they question any "meaningful comparisons of emissions over time" recommended by the GHG Protocol. In particular, the consideration of SSPs seems inevitable to us insofar, as there is an infinity of possible futures, and thus, an infinity of prospective inventories and characterization factors. About inventories, the choice of the IAM generating the scenarios of the probable future is crucial, since it must accurately model the interactions of the biosphere with the production system, linked to investment, innovation, and consumption patterns. In this regard, our literature review highlights the use of REMIND and IMAGE in previous LCA, but more than thirty IAMs exist (Rose 2014; Wilson et al. 2021), with varying characteristics (Wilson et al. 2021), and their adequacy to our sustainable carbon quantification problems must be evaluated. Thereby, the systematic evaluation framework recently proposed by Wilson et al. (2021) must help assess the suitability of these models for our specific purpose. Nevertheless, preparing for the future should not overshadow the importance of accurately evaluating the past, since the reference year for verifying the achievement of emission targets is always in the past. Thus, an inventory and CF database for the past and several sets for probable future scenarios must be developed to allow more robust accounting. Moreover, work must be conducted on how to report and represent emissions over time rather than considering a single emission release for one year of production or one product, taking into account the life cycle relating to organizations' activities and production.

## 6.3. Attributional or consequential approach

There is a long record of vivid discussions about ALCA and CLCA, one part of the community applying only the most common ALCA, another one considering that CLCA is the best way to represent environmental responsibility (Plevin et al. 2014; Weidema et al. 2018), while the vast majority has mixed appreciations or recognizes that each approach has different adequacies for different purposes (Curran et al. 2005; Ekvall et al. 2016; Brander et al. 2019). To our understanding, the different practices can be seen as a continuum between pure attributional and pure consequential approaches, as many approaches are mixed, for instance, consequential models used with attributional background datasets (Weidema 2017). And no approach is completely able to represent any "true environmental impact", this impact being a matter of perspective and calculated with uncertainties. But this continuum of potential quantification rules is a major and transversal challenge that leaks onto all the steps to address carbon neutrality plans. This question must be addressed in the nexus of carbon neutrality pathways when it comes to set system boundaries (Zamagni et al. 2012; Plevin et al. 2014; Weidema et al. 2018), selecting or developing high-quality LCI (Weidema 2017) and electricity LCI (Curran et al. 2005; Vandepaer et al. 2019), approach reduction (Plevin et al. 2014) and neutralizing actions, as well as indirect effects of investments (Sandén and Karlström 2007). In particular, assessing the impact of global or very large-scale system transitions needed to reach carbon neutrality might be highly unreliable without a consequential approach, as this latter considers physical and economic constraints (Weidema et al. 2018) while ALCA does not. On the other hand, CLCA results are not additive, while attributional emissions of all actors in theory aggregate to global emissions (Brander et al. 2019), which may be seen as necessary to calculate trajectories towards net-zero. However, one could argue that the emission reductions to be achieved are so enormous that adopting an accounting system that encourages the most ambitious practices in the short term is what matters the most. We thus invite the LCA community to orientate the ALCA/CLCA discussion in the context of the carbon neutrality emergency to propose the most impactful solutions.

## 6.4. Towards standardization?

Many difficulties encountered by organizations wishing to plan their net-zero transition result from a lack of comprehensive and harmonized guidance. The GHG Protocol documentation, extended by the SBTi, can be seen as the backbone of carbon neutrality guidance. This study highlighted several inconsistencies and limitations to address. The widespread confusion surrounding GHG quantification and carbon neutrality plans can demotivate voluntary corporate actions, justify a lack of dynamism or rigor around carbon neutrality claims, and finally risks the practice of "environmental optimization" mirroring tax optimization, i.e., exploiting methodological or standard loopholes to improve balance sheets.

In response to this, standardizing GHG quantification methods within the framework of carbon neutrality plans seems tempting, or even inevitable to allow comparability needed for action hierarchization and effective "green" investments. But standardization requires massive work and is often part of a timeframe that may seem too long in the face of the climate emergency. Moreover, any standardization of quantification methods reduces the adaptation of quantification to specific or unforeseen cases and may induce protocol flaws (Haya et al. 2020). Despite these limits, we recommend to build such a standardization, based on the work previously accomplished on each of the steps of the MRNC sequence and detailed in this article, as well as plan frequent updates of the standards to include science progress, and adequate databases and accounting tools making the standards and its updates traceable, operational and easily applicable.

## 7. Conclusion

We present the MRNC framework for organizations to plan carbon neutrality pathways, as well as 24 main accounting challenges encountered by organizations and ways forwards to alleviate them by the industrial ecology community. Organizations are generally confused or unaware of the multiplicity of non-harmonized standards, tools and databases developed to tackle net-zero pathways. But a substantial corpus of resources exists and must be referenced, mobilized, updated, extended, completed, and/or homogenized helped by the expertise of the industrial ecology community, before being transmitted for implementation. This implies a bigger involvement from this community within existing climate conglomerates such as the IPCC or the GHG Protocol, or the creation of a new task force dedicated to this work. In particular, high-quality and time-dependent inventories, as well as dynamic GWPs, must be developed and/or refined to consider potential futures, and uncertainties must be accounted for in environmental quantification results. A consensus is still needed on methodological choices in LCA, such as the handling of multifunctional processes, including in the EoL stage, and the role of CLCA and ALCA. A principle for guiding such consensus processes is to give the best chance that the corporate carbon neutrality pathways and planning informed by LCA methods will lead to respecting the 1.5-2°C temperature ceiling of the Paris Agreement.. Finally, while climate change is one of this century's major emergencies, considering other foremost crises such as biodiversity rapid collapse or socioeconomic disasters, burden-shifting must still be controlled in carbon neutrality plans thanks to triple bottom line assessments, using methods such as LCSA, s—or eLCA, and LCC.

**Acknowledgments**

Several ideas presented in this paper originate in a workshop dedicated to the carbon neutrality of organizations prepared at CIRAIG, with the participation of some of its partners. The authors also want to thank Maxime Agez for his advice on EEIO and GHG flows, Anne-France Bolay for her knowledge on CDP and green investments, Dominique Maxime for reviewing the LULUC subsection, Sara Russo-Garrido for her advice on S-LCA, and Titouan Greffe for his references on metal LCIA.

**Disclosure of potential conflicts of interest**

Authors have no conflict of interest to declare.

**Funding**

This study was funded by the International Research Consortium on Life Cycle Assessment and Sustainable Transition at CIRAIG.

**Data transparency**

No dedicated data are part of this study.

**Author contributions**

Conceptualization: ADB, AB; Methodology: ADB, AB; Formal analysis and investigation: ADB; Writing—original draft preparation: ADB, AB; Writing—review and editing: ADB, AB, FS, MM; Funding acquisition: MM;

**FIGURES**

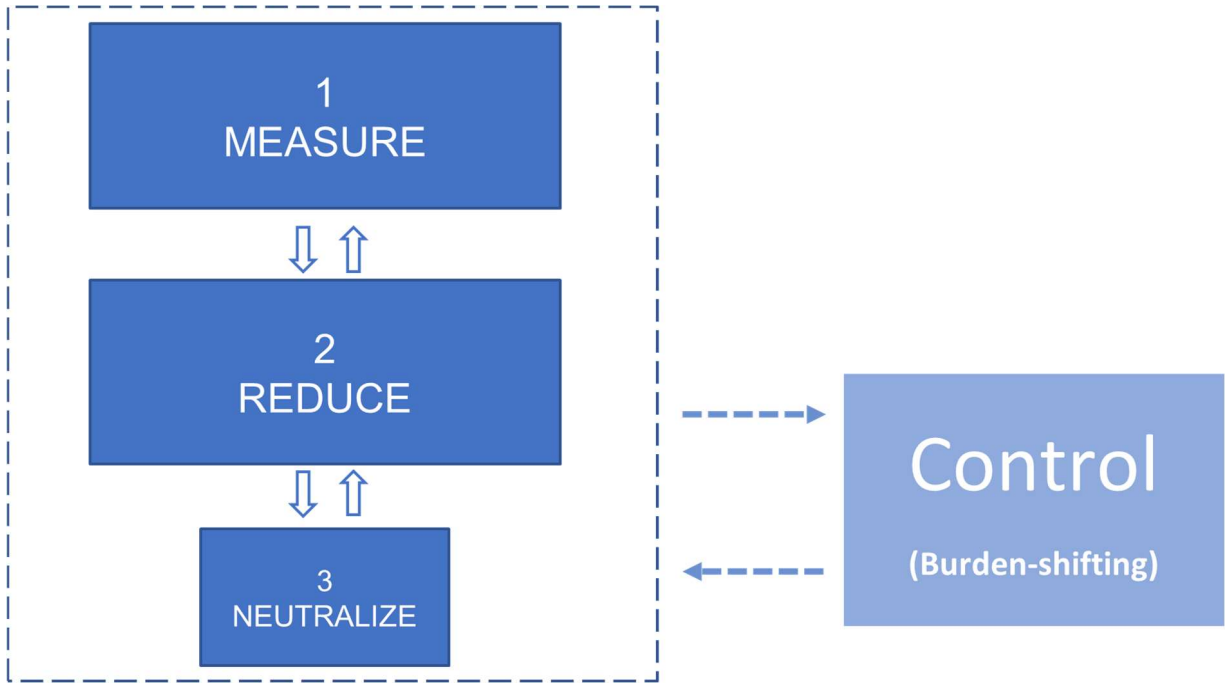

**Figure 1** Generic four-step framework to plan a (sustainable) carbon neutrality pathway

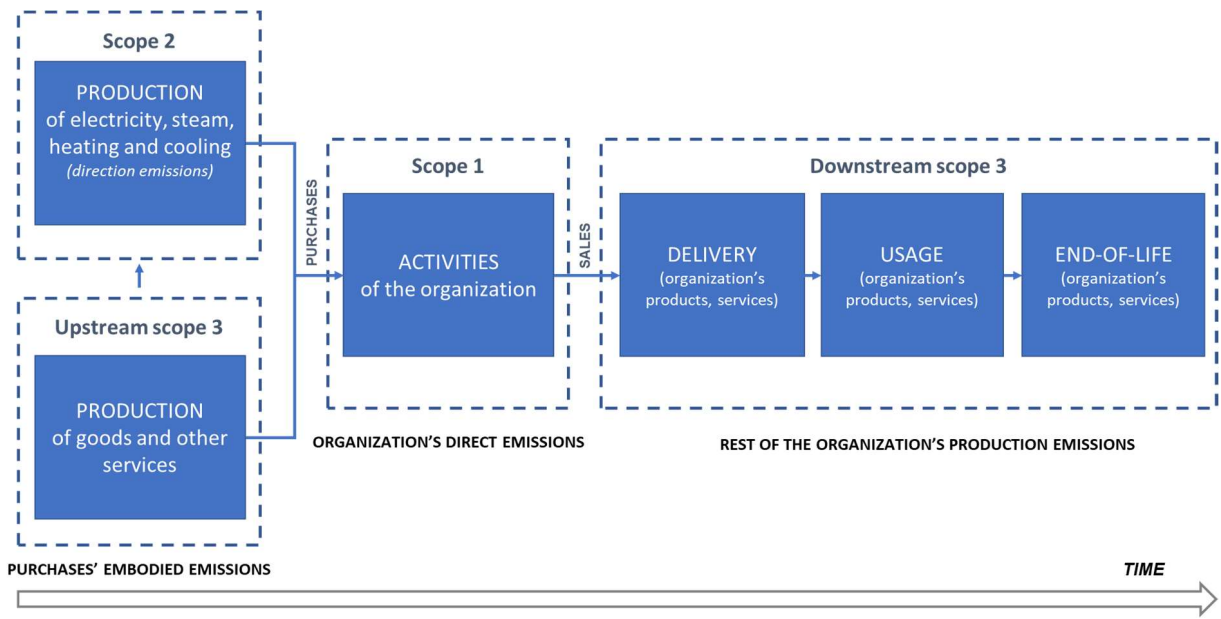

**Figure 2** Reporting scopes from the GHG protocol, reorganized according to an organization's life cycle perspective